# Electrostatic Field Invisibility Cloak

*Chuwen Lan$^{1,2\#}$, Yuping Yang$^{3\#}$, Zhaoxin Geng$^{4\#}$, Bo Li$^{2}$\*, Xianglong Yu$^{1}$, Ji Zhou$^{1}$\**

Dr. C. Lan, Prof. J. Zhou
State Key Laboratory of New Ceramics and Fine Processing, School of Materials Science and Engineering, Tsinghua University, Beijing 100084, China
E-mail: zhouji@mails.tsinghua.edu.cn
Prof. L. Bo, Dr. C. Lan
Advanced Materials Institute, Shenzhen Graduate School, Tsinghua University, Shenzhen, 518000, China
E-mail: libo@mails.tsinghua.edu.cn
Prof. Y.P. Yang
School of Science, Minzu University of China, Beijing, 100081, China.
Prof. Z.X. Geng
School of Information Engineering, Minzu University of China, Beijing, 100081, China.The authors C. Lan，Y.P. Yang and Z. X. Geng contributed equally to this work.

Invisibility cloak is drawing much attention due to its special camouflage when exposed to physical field varing from wave (electromagnetic field, acoustic field, elastic wave, etc.) to scalar field (thermal field, static magnetic field, *dc* electric field and mass diffusion). Here, an electrostatic field invisibility cloak has been theoretically investigated, and experimentally demonstrated for the first time to perfectly hide a certain region from sight without disturbing the external electrostatic field. The desired cloaking effect has been achieved via both scattering cancelling technology and transformation optics (TO). This present work will pave a novel way for manipulating of electrostatic field where would enable a wide range of potential applications and sustainable products made available.

The invisibility cloak refers to a device which has the capability of shielding the object without disturbing the external physical field. Previously, it was believed impossible only made by magicians. Until 2006, an invisibility cloak has been addressed based on the theoretical prediction and experimental demonstration [1-2]. Since then, the intriguing concept has aroused increasing interest, particularly via manipulation of electromagnetic wave [1-8].

Motived by the substantial achievement in electromagnetic wave, considerable efforts are rapidly extended to other waves such as mechanical wave [9], elastic wave [10], and matter waves [11].

Recently, the promising concept of invisibility has also been applied to scalar fields like magnetic field [12, 13], *dc* electric field [14, 15], thermal field [16-20] and mass diffusion [21-24]. However, the possibility for cloaking in the electrostatic field still remains unexplored. Although *dc* electric field is normally used to simulate the electrostatic field, there are obvious differences between them: the former is governed by $J = \sigma E$, whereas, the latter is governed by $D = \varepsilon E$. Hence, the electrostatic field cloak has significant difference from the *dc* electric one (specific before that). On the one hand, since the electrostatic field is widely available ranging from industry, agriculture to daily life, it is therefore expected that the manipulation of the electrostatic field would find potential applications in various fields. For example, the metallic and dielectric materials are usually used to shield or detect the electrostatic field. However, their presence would offer inevitably disturbing effect on the surrounding field and cause physics and engineering experiments to deteriorate. Thus, it is intriguing and promising to develop a cloak sensor to cancel or reduce the disturbing.

Here, we report the first electrostatic field invisibility cloak (EFIC) to shield a certain region without disturbing the external electrostatic field, and this device further experimentally developed can verify the proposed methodology. Our contribution is threefold. Firstly, we have demonstrated the EFIC both theoretically and experimentally for what we believe to be the first time. Secondly, the proposed simple structure can be easily extended to micro-nanoscale and three-dimensional configurations, thereby greatly enhances the practical realization. Finally, the proposed methods, namely, scattering cancelling technology and transformation optics (TO) method, can also be extended to other devices like concentrator, illusion, and rotator.

We start with the EFIC based on cancelling technology method. Figure 1a schematically illustrates the corresponding physical model where a uniform electric field $E$ is applied in $x$-direction with displacement current $D = \varepsilon E$ from high potential to low potential. Assume that an object is placed in the field expected to be cloaked. Here, the idea of the EFIC is to guide the electric field around the object without any distortion as if nothing happens. Inspired by the previous works, the bilayer structure can be easily designed to achieve this goal by directly solving the electrostatic field equation (see *Supporting Material*). The required parameters are given by:

$$\frac{c^2}{b^2} = \frac{(\varepsilon_2 - \varepsilon_1)(\varepsilon_2 + \varepsilon_b)}{(\varepsilon_2 + \varepsilon_1)(\varepsilon_2 - \varepsilon_b)} \tag{1}$$

$\varepsilon_b$, $\varepsilon_1$ and $\varepsilon_2$ are the dielectric constant for the background material, inner layer and outer layer of cloak, respectively. Here, considering that the inner layer of cloak is made of PEC or metal conductor, which has dielectric constant of $\varepsilon_1 \to \infty$ in the static case. Noted that this is different from those in magnetic field, *dc* electric field and thermal field, where $\mu$, $\sigma$, $\kappa$ of the inner layer are set to be 0, respectively. Here, the PEC or metal conductor is chosen due to its excellent electrostatic shielding performance. Furthermore, the material is also widely used as probe in electrostatic measurement. Since $\varepsilon_1 \gg \varepsilon_2$, **Equation 1** can be expressed as

$$\frac{c^2}{b^2} = \frac{-\varepsilon_1(\varepsilon_2 + \varepsilon_b)}{\varepsilon_1(\varepsilon_2 - \varepsilon_b)} = \frac{\varepsilon_2 + \varepsilon_b}{\varepsilon_b - \varepsilon_2} \tag{2}$$

As a result, the required dielectric constant for the outer layer of the cloak can be obtained:

$$\varepsilon_2 = (\frac{c^2 - b^2}{c^2 + b^2})\varepsilon_b \tag{3}$$

**Figure 1**b plots the dependence of relative dielectric constant $\varepsilon_2/\varepsilon_b$ for the outer layer on the radii ratio $c/b$. In our study, the castor oil with the dielectric constant of 4.3 was used as background medium. **Figure 1**b also marks the required geometry parameters when the outer layer is made of air ($\varepsilon = 1.0$) and Teflon ($\varepsilon = 2.1$), respectively. As for practical realization,

the air is chosen as the outer layer, thus the required radii ratio $c/b$ is 1.3. We have chosen the steel layer with $a = 1.3 cm$, $b = 1.5 cm$ as inner layer. Note that other good conductors are also available. The geometry parameters for the air layer can be determined as: $b = 1.5 cm$, $c = 1.95 cm$. To verify the theoretical prediction, numberical simulations were carried out based on Multiphysics Comsol. We also discuss three comparison cases: a) background medium (castor oil); b) castor oil + steel layer (SL); c) castor oil + air layer (AL). In the simulations, the size of the modelling area is $15 \times 15$ cm$^2$, and 1000 V potential was applied to two edges to generate uniform electrostatic field.

Figure 1c-f shows the simulation results, where the electric field distribution and isopotential lines are plotted. Figure 1c corresponds to the case a), Figure 1d and e corresponds to the cases b) and c), respectively, and the Figure 2f provides the result for the designed bilayer cloak. As seen in Figure 1c, a uniform electric field and gradient potential can be generated between the high potential and low potential. Figure 1d and e illustrates that the steel layer repels the isopotential lines and protects the interior from the exterior field while air layer attracts the isopotential lines. For both cases, the isopotential lines and electric field are seriously distorted. In the case of the bilayer cloak, however, the electric field travels around the inner domain as if nothing happens, as depicted in Figure 1f. In contrast, the distortion for electric field and isopotential only occurs in the air layer. Therefore, the inner domain is protected from the external field and thereby a perfect cloak is obtained.

This fabricated bilayer cloak shown in Figure 2a and b is composed of a commercially available steel shell with dimensions of inner radius $a$ = 1.3 cm, outer radius $b$ = 1.5 cm and height $h$ = 5 cm, which is placed in a photosensitive resin container. To construct the bilayer cloak above, the steel shell is further wrapped of a photosensitive resin shell container with the inner radius $b$, outer radius $c$ and height $h$ of 1.5 cm, 1.95 cm and 5 cm, respectively. The thickness of the container wall is 0.45 mm. The shell container is filled with nothing but air, thus a bilayer cloak is achieved. Using simulations, it is found that the use of container would not influence the performance too much. In the experiment, the castor oil is filled in the bigger container and a set of copper plates are used as electrodes. The electrodes were applied with -1000V by electrostatic generator to generate electrostatic field along the $x$-direction. The performance of cloak can be evaluated by obtaining the intensity distribution of the electric field along the line 2.1 cm from the center of bilayer cloak. Clearly, the simulated intensity distribution of electric field for the homogeneous dielectric medium (castor oil) is uniform, which has the value of 6666 V/m. The presence of the steel layer and air layer causes the distortion of electric field, which can be confirmed by the position dependent electric field intensity. As seen in Figure 2c, the electric field intensity near the steel shell is strong and the direction of the electric field has a significant change. The maximum electric field intensity can be determined as 10426 V/m. For the air shell, the electric field intensity decreases near the shell and has a minimum value of 4661 V/m. For the bilayer cloak, the external field is nearly undisturbed and one can achieve uniform electric field intensity with the value about 6660 V/m. Thus, good cloak performance has been obtained. In the measurement, we use the electrostatic instrument to quantify the corresponding electric field. Note that the potential distribution is not measured since it is a challenge to obtain the potential distribution. This is totally different from the *dc* current electric field where potential can be readily measured. The detailed information for electrostatic measuring instrument can be found in *SM*, where the current readings in ampere meter are positive to the electrostatic field intensity detected by the probe. Therefore, one can character the electrostatic field intensity distribution by obtaining the corresponding current in a certain position. The measured results are presented in **Figure 2d**, where the measured current distributions agree well with the simulated electric field intensity distributions, thus validating the feasibility of our scheme. It infers that the deviation can be attributed to the fabrication and measurement.

In addition to scattering cancelling method, the TO theory can also be used to obtain cloak. As shown in Figure 3a, the $x$-axis is a PEC plane connected to the ground. In the transformation, the AOB is stretched to AC'B, while ACB keeps unchanged. Thus, by placing the appropriate materials into the region AC'BCA, one can make the space of AC'BA invisible, then a carpet cloak is achieved. According to the theoretical analysis in *SM*, the required components of dielectric constant tensor in the $x'y'$ system are:

$$\begin{cases} \varepsilon_{x'x'} = \dfrac{(k^2+\tau^2+1)-\sqrt{(k^2+\tau^2-1)^2+4\tau^2}}{2k} \\ \varepsilon_{y'y'} = \dfrac{(k^2+\tau^2+1)+\sqrt{(k^2+\tau^2-1)^2+4\tau^2}}{2k} \end{cases} \quad (4)$$

The rotation between the new and original coordinate system is

$$\theta = \frac{1}{2}\arctan\frac{2\tau}{k^2+\tau^2-1} \quad (5)$$

Clearly, the required material for the carpet cloak is homogenous but anisotropic. To achieve this anisotropy, one can use the metamaterial multilayer structure. Note that one component of the required dielectric constant is larger than background medium and another one is smaller than the background medium. Thus we employ air ($\varepsilon_r = 1.0$) and ultrapure water ($\varepsilon_r = 80.0$) to fabricate such metamaterial. In our study, the geometrical parameters for carpet cloak are: AB = $2a$ =20 cm, OC = $a$ =10 cm, OC' = $0.5a$ = 5cm. As a result, one can obtain that: $\tan\alpha=1$, $\tan\beta=0.5$. The designed metamaterial is presented in Figure 3b, where the filling ratio of the air is about 88%. Simulations were carried out to character the performance of the designed carpet cloak. In the simulations, -1000V potential was biased between the two electrodes to generate nearly uniform electric field. The simulation results for the electric field and potential are shown in the Figure 4a-c. Figure 4a presents that the uniform electric field is generated between the two electrodes, while Figure 4b shows that the presence of cone-shaped PEC causes serious distortion of electric field and potential lines. The simulation result for the carpet cloak is provided in the Figure 4c, where the distortion of the cloak disappears and only occurs in the carpet cloak, revealing a good performance. The performance of the carpet cloak can also be evaluated by the electric field intensity along the dash lines as shown in Figure 3b. The simulated electric field intensity is demonstrated in Figure 4d. The electric field intensity is uniform and can be determined as 6666V/m. When the cone-shape PEC is placed on the ground, the electric field intensity is serious distorted. However, when the PEC is wrapped of carpet cloak, the distortion is cancelled and the electric field intensity becomes uniform again. As schematically shown in Figure 4a-b, the fabricated carpet cloak is a multilayer-groove structure, where the grooves are alternately filled with ultrapure water and air. The measured results are shown in Figure 4f, where the measured current shows good agreement with the simulated electric field intensity, indicating the feasibility of our proposed scheme.

In summary, using scattering cancelling technology and transformation optics (TO) method, we have proposed electrostatic field cloaks that can protect a certain region from the external field as if nothing happens. The proposed cloaks require homogeneous dielectric constant which can be readily obtained with naturally occurring materials, meanwhile the simple structure can be easily extended to micro-nanoscale and three-dimensional configuration, thereby greatly enhances the practical realization and would enable applications like non-destructive detection. More importantly, our concept for manipulation of electrostatic field can also be extended to other devices like concentrator, rotator and illusion, which may find considerable applications in various fields.

**Supporting Information**
Supporting Information is available from the Wiley Online Library or from the author.

**Acknowledgements**

This work was supported by the National Natural Science Foundation of China under Grant Nos. 51032003, 11274198, 51221291 and 61275176, National High Technology Research and Development Program of China under Grant No. 2012AA030403, Beijing Municipal Natural Science Program under Grant No. Z141100004214001, and the Science and

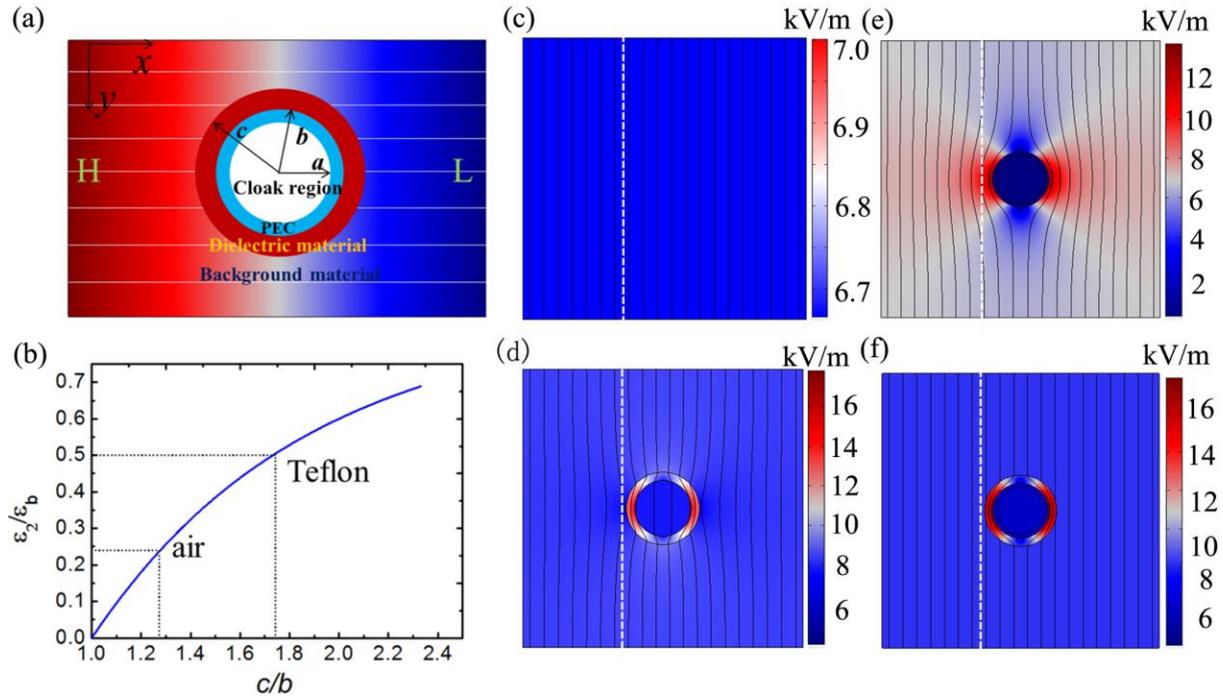

**Figure 1.** (a) The physical model for EFIC. (b) The required relative dielectric constant for EFIC with radii ratio of the outer layer of cloak. The material candidates are marked by dash lines. The simulation electric field intensity and isopotential lines (black lines) for four cases: (c) castor oil. (d) castor oil +SL. (e) castor oil + AL. (f) bilayer cloak. The dotted white lines denote the measuring lines in the experiments.

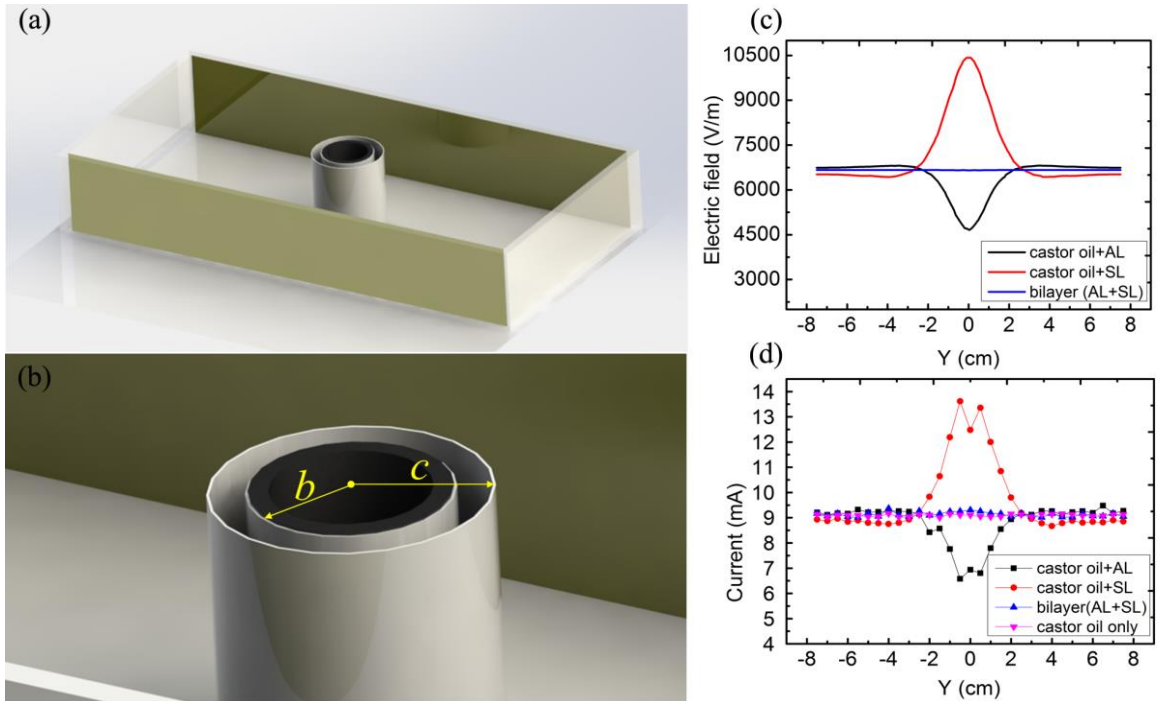

**Figure 2.** Experimental demonstration of EFIC. (a) Schematic illustration for realization of EFIC (b) Schematic illustration of the fabricated sample. (c) Simulated results of electric field intensity for different cases. (f) The measured current in the electrostatic field measurement instrument for corresponding cases, respectively.

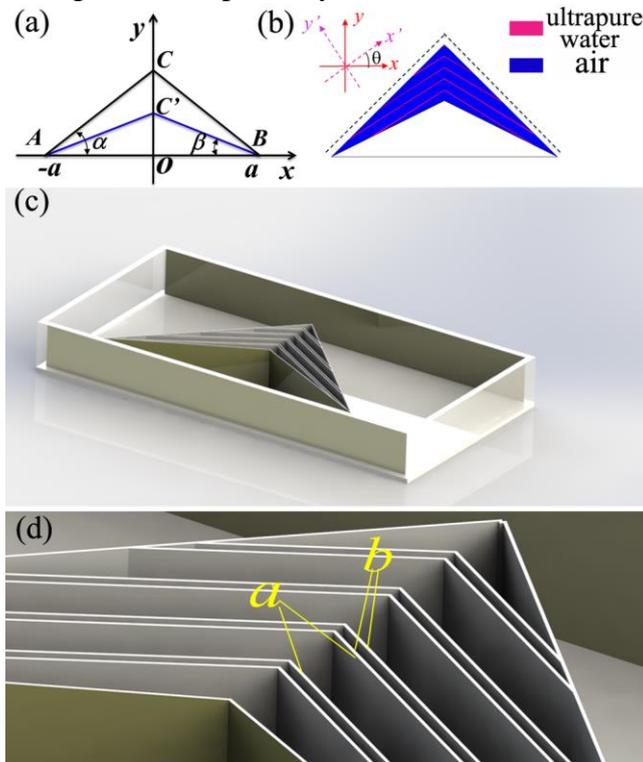

**Figure 3.** (a) The transformation model for the carpet cloak. (b) The designed carpet cloak. The rotation angle between new and old coordinate systems is $\theta$. The black dash lines represent the observation lines. (c) The overall illustration of characterization of the device. (d) zoom, the geometric parameters for the carpet cloak are: $a$= 8.38mm and $b$=1.41mm.

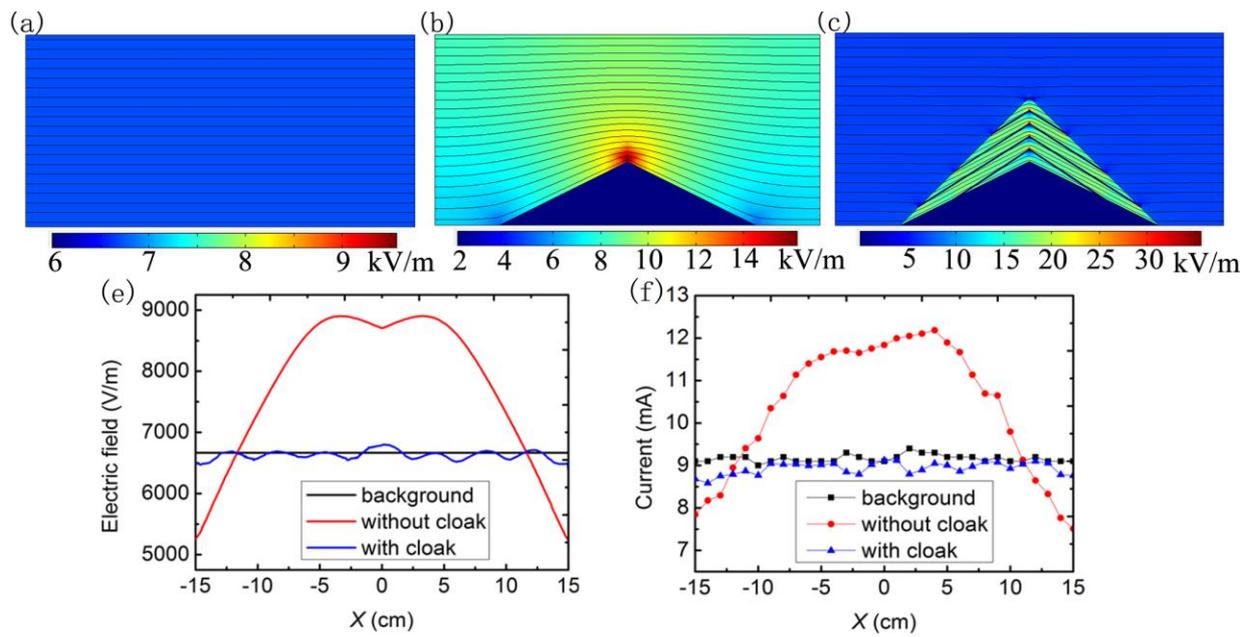

**Figure 4.** The simulated electric field intensity and isopotential lines: (a) background material (b) cone-shape PEC without cloak. (c) cone-shape PEC with carpet cloak. (e) Simulated electric field intensity along the white dash line for the cases. (f) The measured results of current in the electrostatic field measurement instrument for corresponding cases, respectively.

**The electrostatic field invisibility cloak was theoretically and experimentally demonstrated for the first time.** Both scattering cancelling technology and transformation optics method were employed to achieve this goal, which can also be extended to other devices like concentrator, illusion, rotator and etc. This work paves a novel way toward manipulation of electrostatic field which would find considerable potential application.

**Keyword**
Electrostatic field, invisibility cloak, scattering cancelling technology, transformation optics

Chuwen Lan[1,2#], Yuping Yang[3#], Zhaoxin Geng[4#], Bo Li[2]*, Xianglong Yu[1], Ji Zhou[1]*

**Electrostatic Field Invisibility Cloak**

ToC figure

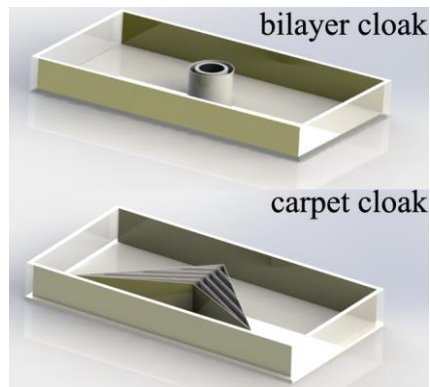

# Supporting Information

## Electrostatic Field Invisibility Cloak

*Chuwen Lan[1,2#], Yuping Yang[3#], Zhaoxin Geng[4#], Bo Li[2]\*, Xianglong Yu[1], Ji Zhou[1]\**

### 1. The theoretical analysis for bilayer cloak

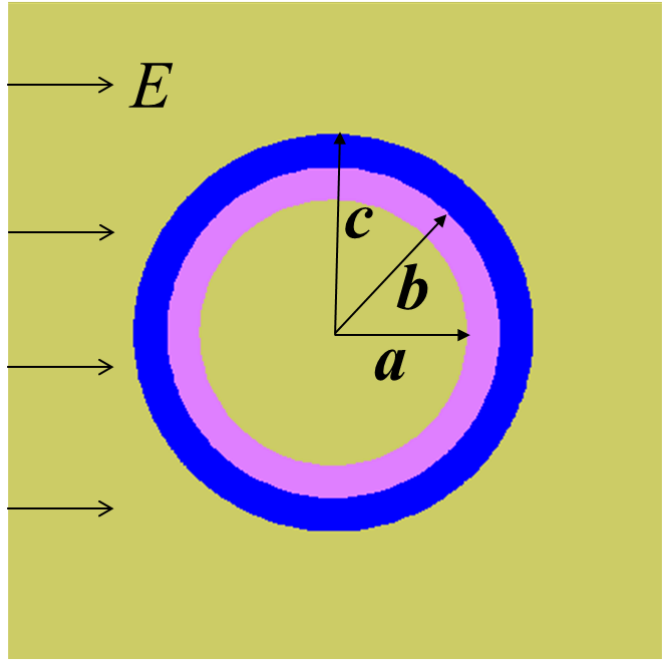

**Figure** S1 The corresponding physical model for bilayer cloak

**Figure S1** schematically illustrates the corresponding physical model where a uniform electric field $E$ is applied in $x$-direction with displacement current $D = \varepsilon E$ from high potential to low potential. In the considered space, the electric potential is governed by the Laplace's equation $\nabla^2 \phi = 0$, which can be expressed as

$$\phi_i = \sum_{m=1}^{\infty} [A_m^i r^m + B_m^i r^{-m}] \cos m\theta \tag{1}$$

where $A_m^i$ and $B_m^i$ ($i$=1, 2, 3) are constants to be determined, and $\phi_i$ represents the potential in different regions: $i$=1 for interior ($r < b$), $i$=2 for cloak shell ($b < r < c$) and $i$=3 for exterior ($r > c$). The dielectric constant for the background material, inner layer and outer layer of cloak shell are $\varepsilon_b$, $\varepsilon_1$ and $\varepsilon_2$, respectively.

Taking into account that $\phi_1$ should be finite when $r \to 0$, one can obtain that $B_1^1 = 0$. In addition, $\phi_3$ should tend to $-E_0 r \cos\theta$ when $r \to \infty$, one only needs to consider $m=1$. Furthermore, the electric potential and the normal component of electric field vector are continuous across the interfaces, one can obtain that

$$\begin{cases} \phi_i |_{r=b,c} = \phi_{i+1} |_{r=b,c} \\ \varepsilon_i \dfrac{\partial \phi_i}{\partial r} |_{r=b,c} = \varepsilon_{i+1} \dfrac{\partial \phi_{i+1}}{\partial r} |_{r=b,c} \end{cases} \quad (2)$$

Here, $\varepsilon_3 = \varepsilon_b$, where $\varepsilon_b$ is the electric dielectric constant of the background. Combining the Equation S1 and S2, one can obtain

$$B_1^3 = E_0 c^2 \frac{\varepsilon_2(M_1 - M_2) - \varepsilon_b(M_1 + M_2)}{\varepsilon_2(M_1 - M_2) + \varepsilon_b(M_1 + M_2)} \quad (3)$$

where $M_1 = c^2(1 + \dfrac{\varepsilon_1}{\varepsilon_2})$, $M_2 = b^2(1 - \dfrac{\varepsilon_1}{\varepsilon_2})$. By making $B_1^3 = 0$, one can obtain

$$\frac{c^2}{b^2} = \frac{(\varepsilon_2 - \varepsilon_1)(\varepsilon_2 + \varepsilon_b)}{(\varepsilon_2 + \varepsilon_1)(\varepsilon_2 - \varepsilon_b)} \quad (4)$$

2. **The theoretical analysis for carpet cloak**

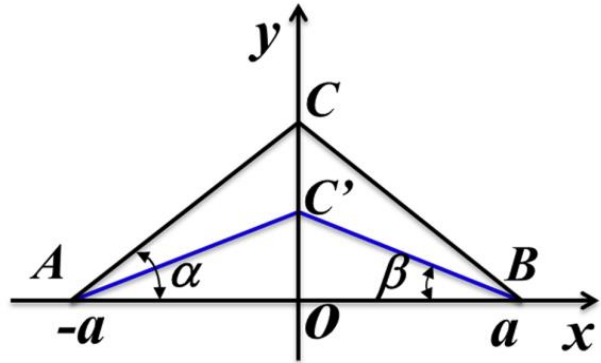

**Figure S2** The transformation model for the carpet cloak, the AOB is stretched to AC'B, while ACB keeps unchanged

In the transformation (see **Figure S2**), the AOB is stretched to AC'B, while ACB keeps unchanged. Thus, by placing the appropriate materials into the region AC'BCA, one can make the space of AC'BA invisible, thus a carpet cloak is achieved. According to the TO theory, one can obtain the required dielectric constant

$$\varepsilon' = \frac{A\varepsilon A^T}{det(A)} \qquad (5)$$

where $A = \frac{\partial(x',y',z')}{\partial(x,y,z)}$ is the Jacobian matrix. Here, the transformation equation is $x'=x$

$$y' = ky + \tau(a-x) \quad z' = z \qquad (6)$$

Then the required dielectric constant can be determined as

$$\varepsilon' = \begin{pmatrix} \frac{1}{k} & \frac{-\tau}{k} & 0 \\ \frac{-\tau}{k} & \frac{(\tau^2+k^2)}{k} & 0 \\ 0 & 0 & \frac{1}{k} \end{pmatrix} \varepsilon \qquad (7)$$

Here $\varepsilon$ is the dielectric constant of the background medium. $k = (\tan\alpha - \tan\beta)/\tan\alpha$, and $\tau = \tan\beta$. For 2D case, only in-plane parameters are considered and they form a symmetric $2\times 2$ matrix. This matrix can be further diagonalized in the $x'y'$ system, where the corresponding components of dielectric constant tensor are

$$\begin{cases} \varepsilon_{x'x'} = \dfrac{(k^2+\tau^2+1) - \sqrt{(k^2+\tau^2-1)^2 + 4\tau^2}}{2k} \\ \varepsilon_{y'y'} = \dfrac{(k^2+\tau^2+1) + \sqrt{(k^2+\tau^2-1)^2 + 4\tau^2}}{2k} \end{cases} \qquad (8)$$

The rotation between the new and original coordinate system is

$$\theta = \frac{1}{2}\arctan\frac{2\tau}{k^2+\tau^2-1} \qquad (9)$$

## 3. Intensity measurement instrument

The circuit of electrostatic field intensity measurement instrument is shown in Figure S3. Electric field induction signal is detected using the characteristic of field-effect tube which has very high input resistance and is very sensitive to electric field induction around it. After switch K is thrown, the source of field-effect tube BG1 and the voltage between drains is lower when there is no electrostatic field around the probe of measurement instrument. There is no current getting through resistance R3, which cuts off BG2. Therefore, collector current of BG2 is zero, ampere meter is zero and the circuit is in the stationary state. When there is electrostatic field around the probe of measurement instrument, the charge begin to accumulate in probe because of electrostatic induction. The bias voltage produced between both ends of resistance R changes the internal resistance of BG1 source and the drain, which results in changes of the whole circuit state. There is current through resistance R3 after breaking over BG2 and the current amplified by BG2 is measured though ampere meter. The probe of measurement instrument can induct different quantity of electric charge in different position of electrostatic field because of different electrostatic field strength. Therefore, there is different collector current of BG2. The electrostatic field intensity can be measured relatively with this method.

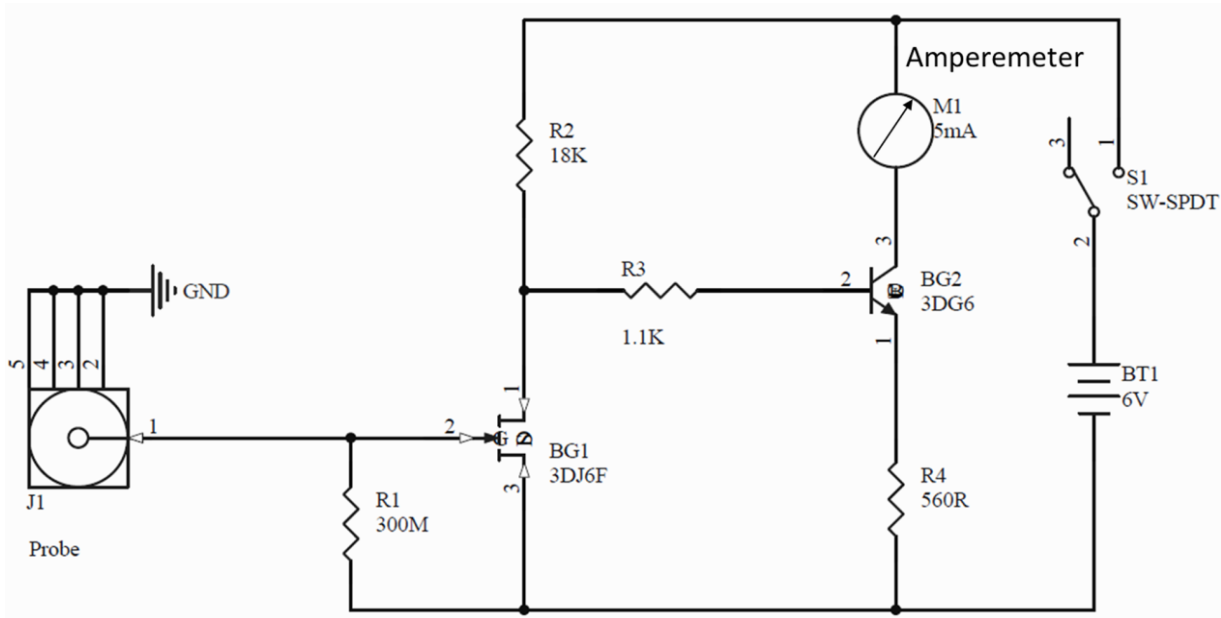

**Figure** S3 The schematic diagram for electrostatic measuring instrument